\RequirePackage{fix-cm}
\documentclass[smallextended]{svjour3}       
\smartqed  
\usepackage{graphicx}
\usepackage{subfigure}
\usepackage{bbding}
\begin{document}

\title{Nonlocal symmetries and exact solutions of variable coefficient AKNS system 
}


\author{Xiangpeng Xin \and  Hanze Liu  \and Yarong Xia 
}


\institute{X. Xin(\Envelope)$^1$  \and H. Liu$^1$ \and Y. Xia$^2$ \at
              1 School of Mathematical Sciences, Liaocheng University, Liaocheng 252059, China \\
              2 School of Information Engineering, Xi'an University, Xi'an 710065, China\\
              \\
              \email{xinxiangpeng2012@gmail.com}           
}

\date{Received: date / Accepted: date}

\maketitle

\begin{abstract}
In this paper, nonlocal symmetries of variable coefficient  Ablowitz-Kaup-Newell-Segur(AKNS) system are discussed for the first time. With lax pair of time-dependent coefficient AKNS system, the nonlocal symmetries are obtained, and they are successfully localized to a Lie point symmetries by introducing a suitable auxiliary dependent variable. Furthermore, using the obtained Lie point symmetries of closed system, we give out two types of symmetry reduction and explicit analytic solutions. For some interesting solutions, the figures are given out to show their dynamic behavior.

\end{abstract}

\section{Introduction}

The theory of Lie groups \cite{Lie1,Ovsiannikov1,Ibragimov1,Bluman1,Liu1,Liu2,Liu3} and their various generalizations have been evolved into one
of the most explosive development of modern mathematics and physics. Nowadays, these theory has been widely applied to diverse fields of mathematics, physics, such as numerical analysis,quantum mechanics, fluid dynamics system, etc.

As a generalization of the symmetry, a lot of studies have been devoted to seeking the general Lie point symmetry. Olver set up so-called nonlocal symmetry\cite{Olver1} theory in the 80s of the last century. Compared with the local symmetry, construction of nonlocal symmetries is a challengeable work and similarity reductions cannot be calculated directly. Recently, some effective techniques to find nonlocal symmetries have been proposed. Bluman et al.\cite{Bluman2,Bluman4} presented many methods to find nonlocal symmetries of partial differential equations (PDEs)by using potential system. Galas \cite{Galas1}obtained the nonlocal Lie-B\"{a}cklund symmetries by introducing the pseudo-potentials. Recently, Lou et al.\cite{Lou3,Hu2,Chen3,Chen4,Xia1} found that Painlev\'{e} analysis can also be applicable in acquiring nonlocal symmetries, also known as residual symmetries.

It's not easy to construct nonlocal symmetries of PDEs, but it is more difficult to construct the nonlocal symmetry of variable coefficient equations. The application of nonlocal symmetry to construct exact solutions of variable coefficient equations is rare. Because the coefficient equations contain the corresponding constant coefficient equations, so the researches have more application value. Here we consider the variable coefficient AKNS system. With the help of lax pair, the nonlocal symmetries of this equation are obtained. Finally, by introducing new variables variables, the nonlocal symmetry is transformed into local symmetry and exact solutions are constructed by using the lie group theory.

This paper is arranged as follows: In Sec.2, the nonlocal symmetries of variable coefficient AKNS system are obtained using the Lax pair. In Sec.3, we transform the nonlocal symmetries into Lie point symmetries by extending original system. The finite symmetry transformation can be obtained by solving the initial value problem. In Sec.4 some symmetry reductions and explicit solutions of the AKNS system are obtained by using the Lie point symmetry of extending system. Finally, some conclusions and discussions are given in Sec.6.

\section{Nonlocal symmetries of variable coefficient AKNS system}

 The time-dependent coefficient AKNS system\cite{Huang1} reads
\begin{equation}\label{ak-1}
\left\{ {\begin{array}{*{20}c}
   {u_t  + \delta (2\alpha vu^2  - \alpha u_{xx} ) = 0,}  \\
   {v_t  - \delta (2\alpha v^2 u - \alpha v_{xx} ) = 0,}  \\
\end{array}} \right.
\end{equation}
where $u = u(x,t)$ and $v = v(x,t)$ are the real functions, $\delta=\delta(t)$ is a real function of $t$. The system(\ref{ak-1}) was obtained via the variable transformation from time-dependent Whitham-Broer-Kaup equations, which is used for the shallow water under the Boussinesq approximation. Lax pair,infinitely-many conservation laws and soliton solutions are given\cite{Huang1}. When $\delta=1, \alpha=i/2$ Eq.(\ref{ak-1}) reduce to the well-known AKNS system, where $i^2=-1$, nonlocal symmetries and explicit solutions for the constant coefficient AKNS system have been obtained\cite{Miao1}. To our knowledge, nonlocal symmetries for Eq.(\ref{ak-1}) have not been obtained and discussed, which will be the goal of this paper.

The corresponding Lax pair has been obtained in\cite{Huang1},
\begin{equation}\label{ak-2}
\begin{array}{l}
 \varphi _x  = U\varphi , \\
 \varphi _t  = V\varphi , \\
 \end{array}
\end{equation}
where
 \begin{center}
 $\varphi  = \left( {\begin{array}{*{20}c}
   {\phi _1 }  \\
   {\phi _2 }  \\
\end{array}} \right),U = \left( {\begin{array}{*{20}c}
   \lambda  & v  \\
   u & { - \lambda }  \\
\end{array}} \right),V = \left( {\begin{array}{*{20}c}
   A & B  \\
   C & { - A}  \\
\end{array}} \right),$
 \end{center}
and
$ A = \alpha \delta uv - 2\lambda ^2 \alpha \delta , B =  - \alpha \delta v_x  - 2\lambda \alpha \delta v,
 C = \alpha \delta u_x  - 2\lambda \alpha \delta u.$

To seek the nonlocal symmetries of variable coefficient AKNS system(\ref{ak-1}), one must solve the following linearized equations,

\begin{equation}\label{ak-3}
\begin{array}{l}
 \sigma _t^1  + 2\alpha vu^2 \sigma ^3  - \alpha u_{xx} \sigma ^3  + 2\alpha \delta u^2 \sigma ^2  + 4\alpha uv\delta \sigma ^1  - \alpha \delta \sigma _{xx}^1  = 0, \\
 \sigma _t^2  - 2\alpha v^2 u\sigma ^3  + \alpha v_{xx} \sigma ^3  - 2\alpha \delta v^2 \sigma ^1  - 4\alpha uv\delta \sigma ^2  + \alpha \delta \sigma _{xx}^2  = 0, \\
\end{array}
\end{equation}
 $\sigma_1,\sigma_2,\sigma_2$ are symmetries of $u,v,\delta$, which means (\ref{ak-1}) is form invariant under the transformations
\begin{equation}\label{ak-3-1}
\begin{array}{l}
u \to u + \epsilon \sigma_1,\\
v \to v + \epsilon \sigma_2,\\
\delta \to \delta + \epsilon \sigma_3,\\
 \end{array}
\end{equation}
with the infinitesimal parameter $\epsilon $.

The symmetry can be written as
\begin{equation}\label{ak-4}
\begin{array}{l}
 \sigma ^1  = \bar X(x,t,u,v,\delta ,\phi _1 ,\phi _2 )u_x  + \bar T(x,t,u,v,\delta ,\phi _1 ,\phi _2 )u_t  - \bar U(x,t,u,v,\delta ,\phi _1 ,\phi _2 ), \\
 \sigma ^2  = \bar X(x,t,u,v,\delta ,\phi _1 ,\phi _2 )v_x  + \bar T(x,t,u,v,\delta ,\phi _1 ,\phi _2 )v_t  - \bar V(x,t,u,v,\delta ,\phi _1 ,\phi _2 ), \\
 \sigma ^3  = \bar T(x,t,u,v,\delta ,\phi _1 ,\phi _2 )\delta _t  - \bar \Delta (x,t,u,v,\delta ,\phi _1 ,\phi _2 ), \\
 \end{array}
\end{equation}

Substituting Eq.(\ref{ak-4}) into Eq.(\ref{ak-3}) and eliminating $u_t,v_t,\phi _{1x},\phi _{1t},\phi _{2x},\phi _{2t}$ in terms of the lax pair(\ref{ak-3}), it yields a system of determining equations for the functions $\bar X,\bar T,\bar U,\bar V,\bar \Delta$ which can be solved by virtue of Maple to give

\begin{equation}\label{ak-5}
\begin{array}{l}
 \bar X(x,t,u,v,\delta ,\phi _1 ,\phi _2 ) = c_1 x + c_2 ,\\
 \bar T(x,t,u,v,\delta ,\phi _1 ,\phi _2 ) = F_1(t), \\
 \bar U(x,t,u,v,\delta ,\phi _1 ,\phi _2 ) = ( - 2c_1  - c_3 )u + c_4 \phi _2^2 , \\
 \bar V(x,t,u,v,\delta ,\phi _1 ,\phi _2 ) = c_3 v + c_4 \phi _1^2 , \\
 \bar \Delta (x,t,u,v,\delta ,\phi _1 ,\phi _2 ) = \delta (2c_1  - \frac{{dF_1(t)}}{{dt}}). \\
 \end{array}
\end{equation}
where $c_i (i = 1, . . . ,4)$ are four arbitrary constants and $F_1(t)$ is arbitrary function of $t$.

\textbf{Remark 1:} It is show that the results(\ref{ak-5}) are local symmetries of variable coefficient AKNS system when $c_4=0$, and they are nonlocal symmetries when $c_4\neq 0$.

Nonlocal symmetries need to be transformed into local ones\cite{Lou3,Hu2} before construct explicit solutions. Hence, we will construct a new system to make the Lie symmetries of the new system contain the nonlocal symmetries(\ref{ak-5}) of the original system.

\section{Localization of the nonlocal symmetry}

For simplicity, letting $c_1  = c_2  = c_3  = 0,c_4  = 1,F_1(t) = 0$ in formula (\ref{ak-5}) i.e.,
\begin{equation}\label{ak-6}
\begin{array}{l}
 \sigma ^1  =  - \phi _2^2 , \\
 \sigma ^2  =  - \phi _1^2 , \\
 \sigma ^3  = 0. \\
\end{array}
\end{equation}

To localize the nonlocal symmetry (\ref{ak-6}), we have to solve the following linearized equations,
\begin{equation}\label{ak-8}
\begin{array}{l}
 \sigma _x^4  - \sigma ^2 \phi _2  - v\sigma ^5  - \lambda \sigma ^4  = 0, \\
 \sigma _x^5  - \sigma ^1 \phi _1  - u\sigma ^4  + \lambda \sigma ^5  = 0, \\
 \sigma _t^4  - \alpha uv\phi _1 \sigma ^3  - \alpha \delta v\phi _1 \sigma ^1  - \alpha \delta u\phi _1 \sigma ^2  - \alpha \delta uv\sigma ^4  + 2\lambda \alpha v\phi _2 \sigma ^3  + 2\lambda \alpha \delta \phi _2 \sigma ^2  \\
  + 2\lambda \alpha \delta v\sigma ^5  + 2\lambda ^2 \alpha \phi _1 \sigma ^3  + 2\lambda ^2 \alpha \delta \sigma ^4  + \alpha \sigma ^3 \phi _2 v_x  + \alpha \delta \phi _2 \sigma _x^2  + \alpha \delta \sigma ^5 v_x  = 0, \\
 \sigma _t^5  + \alpha \delta u\phi _2 \sigma ^2  + \alpha \delta uv\sigma ^5  + \alpha uv\phi _2 \sigma ^3  + \alpha \delta v\sigma ^1 \phi _2  + 2\lambda \alpha u\phi _1 \sigma ^3  + 2\lambda \alpha \delta \phi _1 \sigma ^1  \\
  + 2\lambda \alpha \delta u\sigma ^4  - 2\lambda ^2 \alpha \phi _2 \sigma ^3  - 2\lambda ^2 \alpha \delta \sigma ^5  - \alpha \sigma ^3 \phi _1 u_x  - \alpha \delta \phi _1 \sigma _x^1  - \alpha \delta \sigma ^4 u_x  = 0, \\
 \end{array}
\end{equation}
under the transformations
\begin{equation}\label{ak-3-1}
\begin{array}{l}
\phi _1 \to \phi _1 + \epsilon \sigma_4,\\
\phi _2 \to \phi _2 + \epsilon \sigma_5,\\
f \to f+ \epsilon \sigma_6,\\
 \end{array}
\end{equation}
with the infinitesimal parameter $\epsilon $, and $\sigma_1,\sigma_2,\sigma_3$ given by (\ref{ak-6}). It is not difficult to verify that the solutions of (\ref{ak-8}) have the following forms,
\begin{equation}\label{ak-9}
\sigma ^4  = \phi _1 f,\sigma ^5  = \phi _2 f,
\end{equation}
where $f$ is given by
\begin{equation}\label{ak-9-1}
\begin{array}{l}
 f_x  =  - \phi _1 \phi _2 , \\
 f_t  = \alpha \delta (v\phi _2^2  + 4\lambda \phi _1 \phi _2  - u\phi _1^2 ), \\
 \end{array}
\end{equation}
it is easy to obtain the following result,
\begin{equation}\label{ak-9-2}
\sigma_6=\sigma_f=f^2.
\end{equation}

One can see that the nonlocal symmetry (\ref{ak-6}) in the original space $\left\{ {x,t,u,v,\delta} \right\}$ has been successfully localized to a Lie point symmetry in the enlarged space\\ $\left\{ {x,t,u,v,\delta,\phi1,\phi2,f} \right\}$. It is not difficult to verify that the auxiliary dependent variable $f$ just satisfies the Schwartzian form of the variable coefficient AKNS system
\begin{equation}\label{ak-9-3}
\delta \frac{{\partial C}}{{\partial t}} - \alpha ^2 \delta ^3 \frac{{\partial S}}{{\partial x}} - (8\lambda \alpha \delta ^2  + 3\delta C)\frac{{\partial C}}{{\partial x}} - C\frac{{\partial \delta }}{{\partial t}} = 0,
\end{equation}
where $C = \frac{{\frac{{\partial \phi }}{{\partial t}}}}{{\frac{{\partial \phi }}{{\partial x}}}}$ ,and $S = \frac{{\frac{{\partial ^3 \phi }}{{\partial x^3 }}}}{{\frac{{\partial \phi }}{{\partial x}}}} - \frac{{3\left( {\frac{{\partial ^2 \phi }}{{\partial x^2 }}} \right)^2 }}{{2\left( {\frac{{\partial \phi }}{{\partial x}}} \right)^2 }}$ is the Schwartzian derivative.

After we succeed in making the nonlocal symmetry(\ref{ak-6}) equivalent to Lie point symmetry (\ref{ak-9-1}) of the related prolonged system, the explicit solutions can be constructed naturally by Lie group theory. With the Lie point symmetry(\ref{ak-6}),\\(\ref{ak-9}),(\ref{ak-9-2}), by solving the following initial value problem,
\begin{equation}\label{ak-9-4}
\begin{array}{l}
 \frac{{d\bar u(\epsilon)}}{{d\epsilon }} = -\phi_2^2 ,~~~~\bar u\left| {_{\epsilon  = 0} } \right. = u, \\
 \frac{{d\bar v(\epsilon)}}{{d\epsilon }} =  - \phi_1^2 ,~~~~\bar v\left| {_{\epsilon  = 0} } \right. = v, \\
 \frac{{d\bar \delta(\epsilon)}}{{d\epsilon }} = 0 ,~~~~~~~~\bar \delta \left| {_{\epsilon  = 0} } \right. = \delta, \\
 \frac{{d\bar \phi_1(\epsilon)}}{{d\epsilon }} = \phi_1f,~~~\bar \phi_1\left| {_{\epsilon  = 0} } \right. = \phi_1, \\
 \frac{{d\bar \phi_2(\epsilon)}}{{d\epsilon }} = \phi_2f,~~~\bar \phi_2\left| {_{\epsilon  = 0} } \right. = \phi_2, \\
 \frac{{d\bar f(\epsilon)}}{{d\epsilon }} = f^2 ,~~~~~~~\bar f\left| {_{\epsilon  = 0} } \right. = f, \\
 \end{array}
\end{equation}
where $\epsilon$ is the group parameter.

By solving the initial value problem (\ref{ak-9-4}), we arrive at the symmetry group theorem as follows:

\textbf{Theorem 1.} If $\left\{ {u,v,\delta,\phi _1 ,\phi _2 ,f} \right\}$ is the solution of the prolonged system (\ref{ak-1})(\ref{ak-2})and (\ref{ak-9-1}),with $\lambda = 0$, so is $\left\{ {\bar u,\bar v, \bar \delta, \bar \phi _1 ,\bar \phi _2 ,\bar f}\right\}$

\begin{equation}\label{ak-9-5}
\begin{array}{l}
 \bar u = u + \frac{{\epsilon \phi _2^2 }}{{1 + \epsilon f}},~~~~\bar v = v + \frac{{\epsilon \phi _1^2 }}{{1 + \epsilon f}}, ~~~~\bar \delta = \delta,\\
 \bar \phi _1  = \frac{{\epsilon \phi _1 }}{{1 + \epsilon f}},~~~~~~~~\bar \phi _2  = \frac{{\epsilon \phi _2 }}{{1 + \epsilon f}},~~~~~~~~\bar f = \frac{{\epsilon f}}{{1 + \epsilon f}}, \\
 \end{array}
\end{equation}
with $\epsilon$ is an arbitrary group parameter.

For example, starting from the simple soliton solution of (\ref{ak-1})
\begin{equation}\label{akk-1}
u=-\tanh(2\alpha t-x)-1,v=-\tanh(2\alpha t-x)+1,\delta=1,
\end{equation}
It's not difficult to get the special solutions for the introduced dependent variables from(\ref{ak-2})and (\ref{ak-9-1}),
\begin{equation}\label{akk-2}
 \phi _1  = 1 - \tanh (2\alpha t - x),\phi _2  = 1 + \tanh (2\alpha t - x), f =  - \frac{2}{{1 + e^{4\alpha t - 2x} }}.
\end{equation}

Using theorem 1, it's not hard to verify
\begin{center}
$\begin{array}{l}
 u = \frac{{2(2\varepsilon  - 1)e^{4\alpha t - 2x} }}{{1 - 2\varepsilon  + e^{4\alpha t - 2x} }},v = \frac{2}{{1 - 2\varepsilon  + e^{4\alpha t - 2x} }},\phi _1  = \frac{{2e^{2x} }}{{(2\varepsilon  - 1)e^{2x}  - e^{4\alpha t} }}, \\
 \phi _2  =  - \frac{{2e^{4t\alpha } }}{{(2\varepsilon  - 1)e^{2x}  - e^{4\alpha t} }},f =  - \frac{2}{{1 - 2\varepsilon  + e^{4\alpha t - 2x} }},\delta=1, \\
 \end{array}$
\end{center}
are still the solutions to the system(\ref{ak-1}),(\ref{ak-2})and (\ref{ak-9-1}).

 \textbf{Remark 2:} One can see from the results, the form of the solutions of $u,v$ from the soliton solutions become non-soliton solutions. We can get more solutions by repeating the theorem 1. These solutions can not be obtained by traditional Lie group methods, so they are new exact solutions of system(\ref{ak-1}).

To search for more similarity reductions and exact solutions of Eq.(\ref{ak-1}), we use classical Lie symmetry method to search for similarity reductions of the whole prolonged system and assume the symmetries have the vector form,
\begin{equation}\label{ak-9-6}
 V = X\frac{\partial }{{\partial x}} + T\frac{\partial }{{\partial t}} + U\frac{\partial }{{\partial u}}+ V\frac{\partial }{{\partial v}} + \Delta\frac{\partial }{{\partial \delta}}+ P \frac{\partial }{{\partial p }}  + Q \frac{\partial }{{\partial q }}+ F \frac{\partial }{{\partial f }} ,
\end{equation}
where $X,T,U,\Delta,P,Q,F $ are the functions with respect to ${ x,t,u,\delta,\phi_1,\phi_2,f } $, which means that the closed system is invariant under the transformations
\begin{center}
$(x,t,u,v,\delta ,\phi_1,\phi_2,f )  \to (x + \epsilon X,t + \epsilon T,u + \epsilon U,v + \epsilon V,\delta+ \epsilon \Delta, \phi_1 + \epsilon P,\phi_2 + \epsilon Q,f + \epsilon F)$,
\end{center}
with a small parameter $\epsilon $. Symmetries in the vector form (\ref{ak-9-6}) can be assumed as
\begin{equation}\label{ak-10}
\begin{array}{l}
 \sigma ^1  = X(x,t,u,v,\delta ,\phi _1 ,\phi _2 ,f)u_x  + T(x,t,u,v,\delta ,\phi _1 ,\phi _2 ,f)u_t  - U(x,t,u,v,\delta ,\phi _1 ,\phi _2 ,f), \\
 \sigma ^2  = X(x,t,u,v,\delta ,\phi _1 ,\phi _2 ,f)v_x  + T(x,t,u,v,\delta ,\phi _1 ,\phi _2 ,f)v_t  - V(x,t,u,v,\delta ,\phi _1 ,\phi _2 ,f), \\
 \sigma ^3  = T(x,t,u,v,\delta ,\phi _1 ,\phi _2 ,f)\delta _t  - \Delta (x,t,u,v,\delta ,\phi _1 ,\phi _2 ,f), \\
 \sigma ^4  = X(x,t,u,v,\delta ,\phi _1 ,\phi _2 ,f)\phi _{1x}  + T(x,t,u,v,\delta ,\phi _1 ,\phi _2 ,f)\phi _{1t}  - P_1 (x,t,u,v,\delta ,\phi _1 ,\phi _2 ,f), \\
 \sigma ^5  = X(x,t,u,v,\delta ,\phi _1 ,\phi _2 ,f)\phi _{2x}  + T(x,t,u,v,\delta ,\phi _1 ,\phi _2 ,f)\phi _{2t}  - P_2 (x,t,u,v,\delta ,\phi _1 ,\phi _2 ,f), \\
 \sigma ^6  = X(x,t,u,v,\delta ,\phi _1 ,\phi _2 ,f)f_x  + T(x,t,u,v,\delta ,\phi _1 ,\phi _2 ,f)f_t  - P_3 (x,t,u,v,\delta ,\phi _1 ,\phi _2 ,f), \\
 \end{array}
\end{equation}
$\sigma _i,(i=1,...,6)$ satisfy the linearized equations of the prolonged system, i.e., (\ref{ak-3}),(\ref{ak-8}),and
\begin{equation}\label{ak-10-1}
\begin{array}{l}
 \sigma _x^6  + \sigma _4 \phi _2  + \sigma _5 \phi _1  = 0, \\
 \sigma _t^6  - 4\alpha \lambda \sigma _3 \phi _1 \phi _2  - 4\alpha \lambda \delta \sigma _4 \phi _2  + 2\alpha \delta \sigma _4 \varphi _1 u - 4\alpha \lambda \delta \sigma _5 \phi _1  \\
  - 2\alpha \delta \sigma _5 \phi _2 v + \alpha \sigma _3 \phi _1^2 u - \alpha \sigma _3 \phi _2^2 v + \alpha \delta \sigma _1 \phi _1^2  - \alpha \delta \sigma _2 \phi _2^2  = 0. \\
 \end{array}
\end{equation}

Substituting Eqs.(\ref{ak-10}) into Eqs.(\ref{ak-3}),(\ref{ak-8}),(\ref{ak-10-1}) and eliminating $u_t ,v_t,\phi_{1x},\phi_{1t},\\ \phi_{2x},\phi_{2t},f_x,f_t$ in terms of the closed system, determining equations for the functions $X,T,U,V,\Delta,P,Q,F$ can be obtained, by solving these equations, one can get

\begin{equation}\label{ak-11}
\begin{array}{l}
 X(x,t,u,v,\delta ,\phi _1 ,\phi _2 ,f) = c_1 ,T(x,t,u,v,\delta ,\phi _1 ,\phi _2 ,f) = F_2(t), \\
  U(x,t,u,v,\delta ,\phi _1 ,\phi _2 ,f) = c_2 u + c_3 \phi _2^2 , V(x,t,u,v,\delta ,\phi _1 ,\phi _2 ,f) =  - c_2 v + c_3 \phi _1^2 , \\
 \Delta (x,t,u,v,\delta ,\phi _1 ,\phi _2 ,f) =  - \delta \frac{{dF(t)}}{{dt}},P_1 (x,t,u,v,\delta ,\phi _1 ,\phi _2 ,f) =  - \frac{{\phi _1 }}{2}(c_2  - c_4  + 2c_3 f), \\
 P_2 (x,t,u,v,\delta ,\phi _1 ,\phi _2 ,f) = \frac{{\phi _2 }}{2}(c_2  + c_4  - 2c_3 f), \\
 f(x,t,u,v,\delta ,\phi _1 ,\phi _2 ,f) =  - c_3 f^2  + c_4 f + c_5 , \\
 \end{array}
\end{equation}
where $c_i,(i=1,2,...,5)$ are arbitrary constants, $F_2(t)$ is arbitrary function of $t$.

\textbf{Remark 3: }When $c_1=c_2=c_4=c_5=F_2(t)=0,c_3=-1$, the nonlocal symmetry is just the one expressed by (\ref{ak-5}), and when $c_3=c_4=c_5=0$ the related symmetries only Lie point symmetry of variable coefficient AKNS system.

\section{Symmetry reduction and exact solutions of variable coefficient AKNS system}

In this section, we will give some symmetry reduction and group invariant solutions. Two nontrivial similar reductions under consideration $c_3 \ne 0$ are presented and substantial group invariant solutions are found in the follows.

\textbf{case 1:} $c_5  \ne 0$.

Without loss of generality, we let $c_2=c_4=0,c_1=c_3=1,c_5=k_1,F_2(t)=k_2$, with $k_1,k_2$ are two arbitrary constants. By solving the following characteristic equation,
\begin{equation}\label{ak-13-1}
\frac{{dx}}{1} = \frac{{dt}}{k_2} = \frac{{du}}{{\phi_2^2}} = \frac{{dv}}{{\phi_1^2}} = \frac{{d\delta }}{0 } = \frac{{d\phi _1 }}{{-\phi_1 f }} = \frac{{d\phi _2 }}{{-\phi_2 f }} = \frac{{df}}{{-f^2+k_1 }},
\end{equation}
one can obtain
\begin{equation}\label{ak-14}
\begin{array}{l}
 u = \frac{{\sqrt {k_1 } F_4 (\xi ) - F_3^2 (\xi )\tanh (\Theta )}}{{\sqrt {k_1 } }},v = \frac{{\sqrt {k_1 } F_5 (\xi ) - F_2^2 (\xi )\tanh (\Theta )}}{{\sqrt {k_1 } }} \\
 \phi _1  = F_2 (\xi )\sqrt {\tanh ^2 (\Theta ) - 1} ,\phi _2  = F_3 (\xi )\sqrt {\tanh ^2 (\Theta ) - 1} ,, \\
 f = \sqrt {k_1 } \tanh (\Theta ),\delta  = k_3 . \\
 \end{array}
\end{equation}
where $\Theta  = \sqrt {k_1 } (F_1 (\xi ) + x),\xi  = t - k_2 x$.

Substituting Eqs.(\ref{ak-14})into the prolonged system yields,
\begin{equation}\label{ak-19}
\begin{array}{l}
 F_2  = C e^{\int {\frac{{k_3\alpha k_2^2 F_{1\xi \xi }  + 2k_3\alpha \lambda k_2 F_{1\xi }  - F_{1\xi }  - 2k_3\alpha \lambda }}{{2k_3\alpha k_2 (k_2 F_{1\xi }  - 1)}}d\xi } } ,F_3  = \frac{{k_1  - k_1 k_2 F_{1\xi } }}{{F_2 }}, \\
 F_4  =  - \frac{{ - k_2^2 k_3\alpha F_{1\xi \xi }  + 4k_1 k_2 \lambda k_3\alpha F_{1\xi }  - k_1 F_{1\xi }  - 4k_1 \lambda k_3\alpha }}{{2k_3\alpha F_2^2 }}, \\
 F_5  = \frac{{k_2^2 k_3\alpha F_2^2 F_{1\xi \xi }  + 4k_1 \lambda k_3\alpha F_2^2 F_{1\xi }  - 4\lambda k_3\alpha F_2^2  + F_2^2 F_{1\xi } }}{{2k_3\alpha k_2^2 F_{1\xi }^2  - 4k_3\alpha k_1 k_2 F_{1\xi }  + 2k_3\alpha k_1 }}, \\
 \end{array}
\end{equation}
where $C$ is arbitrary constant. One can see that through the Eqs.(\ref{ak-14})and (\ref{ak-19}), if we know the form of $F_1(\xi)$, then $u,v$ can be obtained directly. And we known that auxiliary dependent variable $f$ satisfies the Schwartzian form, by substituting $f = \sqrt {k_1 } \tanh (\Theta )$ into (\ref{ak-9-3}), one can get,
\begin{equation}
\begin{array}{l}
 \alpha ^2 k_3^2 k_2^4 (2k_2 F - k_2^2 F^2  - 1)F_{\xi \xi \xi }  - 3k_3^2 \alpha ^2 k_2^6 F_\xi ^3  + [4\alpha ^2 k_3^2 k_2^5 (k_2 F - 1)F_{\xi \xi }  \\
  + 4k_1 k_3^2 \alpha ^2 k_2^4 F^2 (k_2^2 F^2  - 4k_2 F + 6) + (2k_2  - 16k_1 k_2^3 k_3^2 \alpha ^2  - 4k_3\alpha \lambda k_2^2 )F \\
  + 4k_1 k_3^2 \alpha ^2  + 4k_3\lambda \alpha k_2  + 1]F_\xi   = 0, \\
 \end{array}
\end{equation}
where $F_{1\xi }=F(\xi)=F$.

It is not difficult to verify that the above equation is equivalent to the following elliptic equation,
\begin{equation}\label{ak-19-1}
F_\xi   = \frac{1}{{k_3\alpha k_2^3 }}\sqrt {A_0  + A_1 F + A_2 F^2  + A_3 F^3  + A_4 F^4 }£¬
\end{equation}
where
\begin{center}
$\begin{array}{l}
 A_0  = 2k_3C_1 \alpha ^2 k_2^5  + 2k_3^2 C_2 \alpha ^2 k_2^5  + 4C\alpha \lambda k_2  - 1, \\
 A_1  =  - (4k_3^2 C_1 \alpha ^2 k_2^6  + 6k_3^2 C_2 \alpha ^2 k_2^6  + 4k_3\alpha \lambda k_2^2  - 2k_2 ), \\
 A_2  = 2k_3^2 C_1 \alpha ^2 k_2^7  + 6k_3^2 C_2 \alpha ^2 k_2^7  + 4k_3^2 \alpha ^2 k_1 k_2^4 , \\
 A_3  =  - 2k_3^2 C_2 \alpha ^2 k_2^8  - 8k_3^2 \alpha ^2 k_1 k_2^5 , \\
 A_4  = 4k_3^2 \alpha ^2 k_1 k_2^6 . \\
 \end{array}$
\end{center}
$C_1,C_2$ are arbitrary constants.

It is know that the general solution of Eq.(\ref{ak-19-1}) can be written in terms of Jacobi elliptic functions. Hence, expression of solution (\ref{ak-14}) reflects the
wave interaction between the soliton and the Elliptic function periodic wave. A simple solution of Eq.(\ref{ak-19-1}) is given as,
\begin{equation}\label{ak-19-2}
F = b_0  + b_1 sn(\xi ,n),
\end{equation}
substituting Eq.(\ref{ak-19-2}) into Eq.(\ref{ak-19-1}) yields
\begin{equation}\label{ak-19-3}
b_0  = 2\alpha \lambda k_3,b_1  = 8k_3^2 \alpha ^2 \lambda ^3 ,k_1  = \frac{{n^2 }}{{256k_3^4 \alpha ^4 \lambda ^6 }},k_2  = \frac{1}{{2\lambda \alpha k_3}},
\end{equation}
with $k_3,\lambda ,\alpha  \in R, 0 \le n \le 1$.

Substituting Eqs.(\ref{ak-19-3}),(\ref{ak-19-2}) and $F_{1\xi}=F$ into Eq.(\ref{ak-19}), one can obtain the solutions of $u,v$. Because the expression is too prolix, it is omitted here. In order to study the properties of these solutions of AKNS system, we give some pictures of $u,v$ as following,


In Fig.1, we plot the interaction solutions between solitary waves and cnoidal waves expressed by (\ref{ak-19}) with parameters $C=5,C_1=2,k_1=0.18,k_2=10,\lambda=0.1,\alpha=1,n=0.1$.

We can see that the component $u$ exhibits a soliton propagates on a Elliptic sine function wave background. In Fig.1, the first picture(a) shows that the height of the soliton is approximately 0.03 at $t=-10$. With the development of time, soliton produces elastic collisions with other waves, and the height increases continuously. Picture(e) shows that soliton is roughly in line with its adjacent wave at $t=14$. After the collision, the soliton reverts to the original height and continues to collide with the adjacent waves see the pictures($f \to i$). The corresponding 3d image is given below, exhibits a soliton propagating on period waves background. As one can see from the expression(\ref{ak-14}), $u,v$ possess similar form, so there is no more detailed discussion here.

In order to study the properties of the solutions, we draw the corresponding 3-D images using Maple software,(see Fig.2) and the parameters used in the figures are selected same as Fig.1.

In fact, it is of interest to study these types of solutions, for example, in describing localized states in optically refractive index gratings. In the ocean, there are some typical nonlinear waves such as the solitary waves and the cnoidal periodic waves.

\textbf{case 2:} $c_5 =0$.

We let $c_1  = k_1 ,c_2  = 2k_2 ,c_3  = k_3 ,c_4  = c_5  = 0,F_1 (t) = 1$, with $k_1,k_2,k_3$ are arbitrary constants. By solving the following characteristic equation,

\begin{equation}\label{akkk-1}
\begin{array}{l}
 \frac{{dx}}{{k_1 }} = \frac{{dt}}{1} = \frac{{du}}{{k_3 \phi _2^2  + 2k_2 u}} = \frac{{dv}}{{k_3 \phi _1^2  + 2k_2 v}} = \frac{{d\delta }}{0} \\
  = \frac{{d\phi _1 }}{{ - \phi _1 (k_3 f + k_2 )}}= \frac{{d\phi _2 }}{{ - \phi _2 (k_3 f - k_2 )}} = \frac{{df}}{{ - k_3 f^2 }}, \\
 \end{array}
\end{equation}
one can obtain the following results,

\begin{equation}\label{akkk-2}
\begin{array}{l}
 u = e^{2k_2 t} (\tilde F_4 (\varsigma ) - \frac{{\tilde F_3^2 (\varsigma )}}{{\tilde F_1 (\varsigma ) + k_3 t}}),v = e^{ - 2k_2 t} (\tilde F_5 (\varsigma ) - \frac{{\tilde F_2^2 (\varsigma )}}{{\tilde F_1 (\varsigma ) + k_3 t}}), \\
 \phi _1  = \frac{{e^{ - k_2 t} \tilde F_2 (\varsigma )}}{{\tilde F_1 (\varsigma ) + k_3 t}},\phi _2  = \frac{{e^{k_2 t} \tilde F_3 (\varsigma )}}{{\tilde F_1 (\varsigma ) + k_3 t}},f = \frac{1}{{\tilde F_1 (\varsigma ) + k_3 t}},\delta  = \tilde C. \\
 \end{array}
\end{equation}
where $\varsigma  = x - k_1 t$, $\tilde C$ is a arbitrary constant.

Substituting Eqs.(\ref{akkk-2})into the prolonged system yields,
\begin{equation}\label{akkk-3}
\begin{array}{l}
 \tilde F_2  = \tilde C_1 e^{\int { - \lambda  + \frac{{\tilde F_{1\varsigma \varsigma } }}{{\tilde F_{1\varsigma } }} + \frac{{k_1 }}{{2\tilde C\alpha }} - \frac{{k_3 }}{{2\tilde C\alpha _2 \tilde F_{1\varsigma } }}d\varsigma } } ,\tilde F_3  = \frac{{\tilde F_{1\varsigma } }}{{\tilde F_2 }}, \\
 \tilde F_4  =  - \frac{{ - \tilde C\alpha \tilde F_{1\varsigma \varsigma }  - 4\lambda \tilde C\alpha \tilde F_{1\varsigma }  + k_1 \tilde F_{1\varsigma }  - k_3 }}{{2C\alpha \tilde F_2^2 }}, \\
 \tilde F_5  = \frac{{\tilde C\alpha \tilde F_2^2 F_{1\varsigma \varsigma }  - 4\lambda \tilde C\alpha \tilde F_2^2 \tilde F_{1\varsigma }  + k_1 \tilde F_2^2 \tilde F_{1\varsigma }  - k_3 \tilde F_2^2 }}{{2\tilde C\alpha \tilde F_{1\varsigma }^2 }}, \\
 \end{array}
\end{equation}
where $\tilde C_1$ is a arbitrary constant and $F=F(\varsigma)=F_{1\varsigma}$ satisfies the following equation
\begin{equation}\label{akkk-4}
\tilde C^2 \alpha ^2 (\tilde F^2 \tilde F_{\varsigma \varsigma \varsigma }  + 3\tilde F_\varsigma ^3 ) - (4\tilde C^2 \alpha ^2 \tilde F\tilde F_{\varsigma \varsigma }  + 4\tilde C\alpha \lambda k_3 \tilde F - 2k_1 k_3 \tilde F + 3k_3^2 )\tilde F_\varsigma   = 0,
\end{equation}
the equation(\ref{akkk-4}) is equivalent to the following elliptic equation,

\begin{equation}\label{akkk-5}
\tilde F_\varsigma   = \frac{{\sqrt { - 2\tilde C^2 \alpha ^2 \tilde C_1 \tilde C_2 \tilde F^3  + 2\tilde C^2 \alpha ^2 \tilde C_1 \tilde F^2  + (4\tilde Ck_3 \alpha \lambda  - 2k_1 k_3 )\tilde F + k_3^2 } }}{{\tilde C\alpha }}.
\end{equation}

To solve the equation(\ref{akkk-5}), we assume a solution with the following form,

\begin{equation}\label{akkk-6}
\tilde F = \frac{1}{{l_0  + l_1 sn(\varsigma ,m)}},
\end{equation}
substituting Eq.(\ref{akkk-6}) into Eq.(\ref{akkk-4}) yields the following eight sets of solutions,

\begin{equation}\label{akkk-7}
\begin{array}{l}
 \{ k_1  =  \pm 2\tilde Cm\alpha  + 2\tilde C\alpha \lambda ,k_3  = \frac{{\tilde Cm\alpha }}{{l_1 }},l_0  = l_1 \} , \\
 \{ k_1  = 2\tilde C\alpha \lambda  \pm 2\tilde C\alpha ,k_3  =  \pm \frac{{\tilde C\alpha }}{{l_0 }},l_1  =  \pm l_0 m\} , \\
 \{ k_1  =  \pm 2\tilde Cm\alpha  + 2\tilde C\alpha \lambda ,k_3  =  - \frac{{\tilde Cm\alpha }}{{l_1 }},l_0  =  \mp l_1 \} , \\
 \{ k_1  = 2\tilde C\alpha \lambda  \pm 2\tilde C\alpha ,k_3  =  \pm \frac{{\tilde C\alpha }}{{l_0 }},l_1  =  \mp l_0 m\} , \\
 \end{array}
\end{equation}

\textbf{Remark 4:} Substituting Eqs.(\ref{akkk-7}),(\ref{akkk-6}) and (\ref{akkk-3}) into Eq.(\ref{akkk-2}) yields the exact solutions of variable coefficient AKNS system(\ref{ak-1}). It can be known from the expression (\ref{akkk-2}) that $u,v$ are rational function form solutions. If take $k_2=0$, then solutions are transformed into elliptic function solutions.

\section{ Summary and Discussion}

In this paper, we have studied nonlocal symmetries and exact solutions of the variable coefficient AKNS system. and obtained the following results for the first time.

First of all, starting from the known Lax pairs of the variable coefficient AKNS system, nonlocal symmetries are derived directly. In order to take advantage of the nonlocal symmetries, an auxiliary variable is introduced. Then, the primary nonlocal symmetry is equivalent to a Lie point symmetry of a prolonged system. Applying the Lie group theorem to these local symmetries, the corresponding group invariant solutions are derived.

Secondly, several classes of exact solutions are provided in the paper, including some special forms of exact solutions. For example, exact interaction solutions
among solitons and other complicated waves including periodic waves. This kind of solution can be easily applicable to the analysis of physically interesting processes.

Using Lax pairs to search for nonlocal symmetries and exact solutions of variable coefficient integrable models are both of considerable interest. However, there is not a universal way to estimate what kind of nonlocal symmetries can be localized to the Lie point symmetries. It is more difficult to obtain similar conclusions for variable coefficient differential equations. Above topics will be discussed in the future series research works.

\small{

\end{document}